\documentclass[pre,floats,twocolumn,showpacs,superscriptaddress]{revtex4-1}

\usepackage{graphicx}
\usepackage{amssymb}
\usepackage{color}
\usepackage{amsthm}
\usepackage{ulem}
\usepackage{url}
\usepackage{dcolumn,bm,fleqn,epic,eepic,float}
\usepackage{subfigure}

\begin{document}

\title{Emergence of network features from multiplexity}

\author{Alessio Cardillo}
\affiliation{Institute for Biocomputation and Physics of Complex Systems (BIFI), Universidad de Zaragoza, E-50018 Zaragoza, Spain.}
\affiliation{Departamento de F\'{\i}sica de Materia Condensada, Universidad de Zaragoza, E-50009 Zaragoza, Spain.}

\author{Jes\'us G\'omez-Garde\~{n}es}
\email{gardenes@gmail.com}
\affiliation{Institute for Biocomputation and Physics of Complex Systems (BIFI), Universidad de Zaragoza, E-50018 Zaragoza, Spain.}
\affiliation{Departamento de F\'{\i}sica de Materia Condensada, Universidad de Zaragoza, E-50009 Zaragoza, Spain.}

\author{Massimiliano Zanin}
\affiliation{Center for Biomedical Technology, Universidad Polit\'ecnica de Madrid, E-28223 Pozuelo de Alarc\'on, Madrid, Spain.}
\affiliation{Innaxis Foundation \& Research Institute, Jos\'e Ortega y Gasset 20, 28006, Madrid, Spain.}
\affiliation{Faculdade de Ci\^encias e Tecnologia, Departamento de Engenharia Electrot\'ecnica, Universidade Nova de Lisboa, Portugal.}

\author{Miguel Romance}
\affiliation{Center for Biomedical Technology, Universidad Polit\'ecnica de Madrid, E-28223 Pozuelo de Alarc\'on, Madrid, Spain.}
\affiliation{Departmento de Matem\'atica Aplicada, Universidad Rey Juan Carlos, E-28933 M\'ostoles, Madrid, Spain.}

\author{David Papo}
\affiliation{Center for Biomedical Technology, Universidad Polit\'ecnica de Madrid, E-28223 Pozuelo de Alarc\'on, Madrid, Spain.}

\author{Francisco del Pozo}
\affiliation{Center for Biomedical Technology, Universidad Polit\'ecnica de Madrid, E-28223 Pozuelo de Alarc\'on, Madrid, Spain.}

\author{Stefano Boccaletti}
\affiliation{Center for Biomedical Technology, Universidad Polit\'ecnica de Madrid, E-28223 Pozuelo de Alarc\'on, Madrid, Spain.}

\begin{abstract}
Many biological and man-made networked systems are characterized by the simultaneous presence of different sub-networks organized in separate layers, with links and nodes of qualitatively different types. While during the past few years theoretical studies have examined a variety of structural features of complex networks, the outstanding question is whether such features are characterizing all single layers, or rather emerge as a result of coarse-graining, i.e. when going from the multilayered to the aggregate network representation. Here we address this issue with the help of real data. We analyze the structural properties of an intrinsically multilayered real network, the European Air Transportation Multiplex Network in which each commercial airline defines a network layer. We examine how several structural measures evolve as layers are progressively merged together. In particular, we discuss how the topology of each layer affects the emergence of structural properties in the aggregate network.
\end{abstract}

\maketitle


In the past fifteen years, network theory\cite{rev:albert,rev:newman,boccaletti} has successfully characterized the interaction among the constituents of a variety of complex systems\cite{watts, strogatz}, ranging from biological\cite{alon} to technological\cite{barabasi}, and social\cite{wasserman} systems. However, up until recently, attention was almost exclusively given to networks in which all components were treated on equivalent footing, while neglecting all the extra information about the temporal- or context-related properties of the interactions under study. Only in the last three years, taking advantage of the enhanced resolution in real data sets, network scientists have directed their attention to the multiplex character of real-world systems, and explicitly considered the time-varying \cite{onnela,tang,bianconi,catuto,mucha10,szell} and multi-layered \cite{kurant,Buldyrev10,parshani,Nphys,lee,souza,brummitt, reinares, gomez,criado,ronhovde09,ronhovde11} nature of networks.

A paradigmatic example of intrinsically multiplex system is represented by the Air Transportation Network (ATN). The ATNs have undergone a very significant growth during the last decades, giving rise to the dense and redundant system we know nowadays\cite{EUROCONTROL10}. In the ATN, nodes represent airports, while links stand for direct flights between two airports. On the other hand, each commercial airline corresponds to a different layer, containing all the connections  operated by the same company. While a considerable effort has recently been devoted to the characterization of the structural properties\cite{guimera,colizza,wuellner} of ATNs and their role in the dynamical processes taking place on them\cite{colizza2,colizza3,lacasa,meloni}, their multiplex nature has remained almost unexplored.

When studying systems that can be represented as a graph made of diverse relationships (layers) between its constituents, an important question, typical of complex systems analysis, arises: can the topological properties of the whole system be traced to those of its layers or do they {\it emerge} from the simultaneous presence of multiple layers? Emergence is said to happen when the focus is switched from one scale to a coarser level of description. This question can be addressed by comparing the most usual structural properties of the multiple layers composing a network \cite{newman} and their analogue in the aggregate representation of the network, in which the layer structure is disregarded.

To address the above question we resort to the European ATN data set. Taking advantage of the high-resolution of these data, comprising a number of airlines (layers) operating in Europe during the year $2011$, we succeed to extract the multiplex character of the system, and we investigate how the structural properties usually observed in the ATN are here emerging as a result of progressive layer merging.  To this end, we quantify various topological measures, such as the degree distribution, the clustering coefficient or the presence of rich-club effect, in networks obtained by merging together a growing number of layers, from the lowest level of resolution of a single layer, up to the fully {\it aggregate} network. In addition, we compare two different types of layers, those corresponding to major (national) airlines and those labeled as low-cost companies. We analyze their structural differences, and their different contribution to the properties of the global ATN.

\section*{Results}

\begin{figure*}[ht!]
\centering
\includegraphics[width=\textwidth]{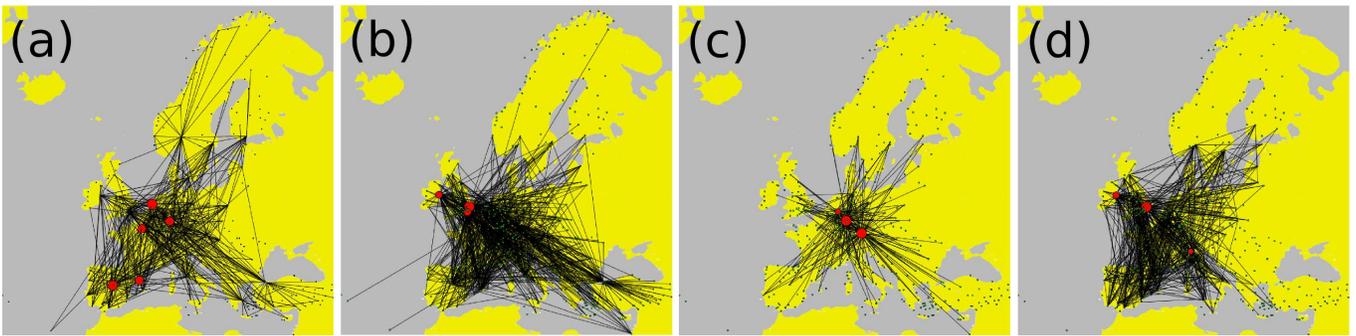}
\caption{{Visual representation of the ATN. From left to right: the aggregate network of all the layers in which only links belonging to more than one layer are displayed. The same network but in which we display those links which belongs to only one layer and connecting at least one node with degree greater than or equal to 75. An example of ATN network of a major airline and, finally, the network of a low-fare (low-cost) airline. In each network, the airport with the highest degree are highlighted.}}
\label{fig:air-net}
\end{figure*}

The European ATN can be represented as a graph composed of $M=37$ different layers each representing a different European airline (see Methods for details). Each layer $m$ has the same number of nodes, $N$, as all European airports are represented in each layer. Furthermore, the data set allows extracting two main subsets, comprising all major, and low-cost airlines, with $18$ and $10$ layers respectively (See Fig.~1).
In particular, panels {\bf (a)} and {\bf (b)} display the structure of the aggregate network focusing first on its {\it redundancy} by sketching those links belonging to more than one layer and on its {\it unicity} by reporting those links that only exist in a specific layer. Panels {\bf (c)} and {\bf (d)} show, instead, the single-layer ATN corresponding to a given major and low-cost airlines, respectively. In each of the panels we highlighted the nodes with the three highest number of connections.

\subsection{Topological measures}

To characterize the structural properties of both the aggregate ATN and its layers, we consider several features widely used in network literature\cite{newman}, i.e. cumulative degree distribution $P_>(k)$, clustering coefficient $C$, size of the giant component $S$, average path length $L$ and Rich-club coefficient $R$. We briefly describe below the specific meaning of each of these measures in our context. The interested reader will find a complete description of all those quantities in the Methods section.

\begin{itemize}
\item The cumulative degree distribution $P_>(k)$, gives the probability of finding a node with a number of connections (or degree) equal or greater than $k$. The degree distribution is a powerful tool which allows understanding both structural and dynamical characteristics of a system as, for instance, its tolerance to attacks or failures\cite{cohen_2000, cohen_2001} so it represents a cornerstone in the characterization of critical infrastructures, such as the ATN.

\item The average path length $\langle L \rangle$, measures the average number of hops one has to make to go from a node to another. In the context of ATNs, it indicates the {\it average number of flights} a passenger has to take to go from his/her origin to his/her destination. However, if the system is not connected, this quantity diverges and it is preferable to restrict attention to the giant (largest) component of the system (see below).

\item The clustering coefficient $C$, measures the probability, $C\in[0,1]$, that two nodes with a common neighbor are connected together. $C$ is a typical measure in systems made of social acquaintances\cite{wasserman}, but in our case it is useful to estimate the density of triangular motifs (denoting the possibility of performing round trips of length $3$).

\item The size of the giant component\cite{stauffer} $S$, denotes the largest fraction of overall nodes such that any pair of them is connected through a path of finite length. In our case, it estimates the largest {\it coverage} that a given airline (or a combination of them) provides in terms of the available destinations that a passenger can reach from an origin inside the giant component.

\item The Rich-club coefficient\cite{zhou} $R$, measures the tendency of highly connected nodes, {\it i.e.} the {\it hubs}, to be connected among themselves. To measure it, one has to compute the abundance of links, $\phi(k)$, among nodes with a number of connections equal or greater than a certain value $k$, and the maximum possible number of links among those nodes, $\phi(k)_{max}$. Then, the ratio between these two quantities gives the relative abundance of links among nodes with at least $k$ connections. Finally, $R(k)$ is given by the ratio between the abundance of links in the real case $\phi(k)/\phi(k)_{max}$ and the same quantity calculated in a proper randomized version of the original network. Colizza {\it et al.} \cite{colizza} measured $R$ for the ATN, and found that world air transportation network displays indeed a Rich-club effect, {\it i.e.} for large values of $k$ the value of $R(k)$ is larger than $1$.
\end{itemize}

\subsection{Emergence of topological properties of the European ATN}

\begin{figure*}[ht!]
\centering
\includegraphics[width=\textwidth]{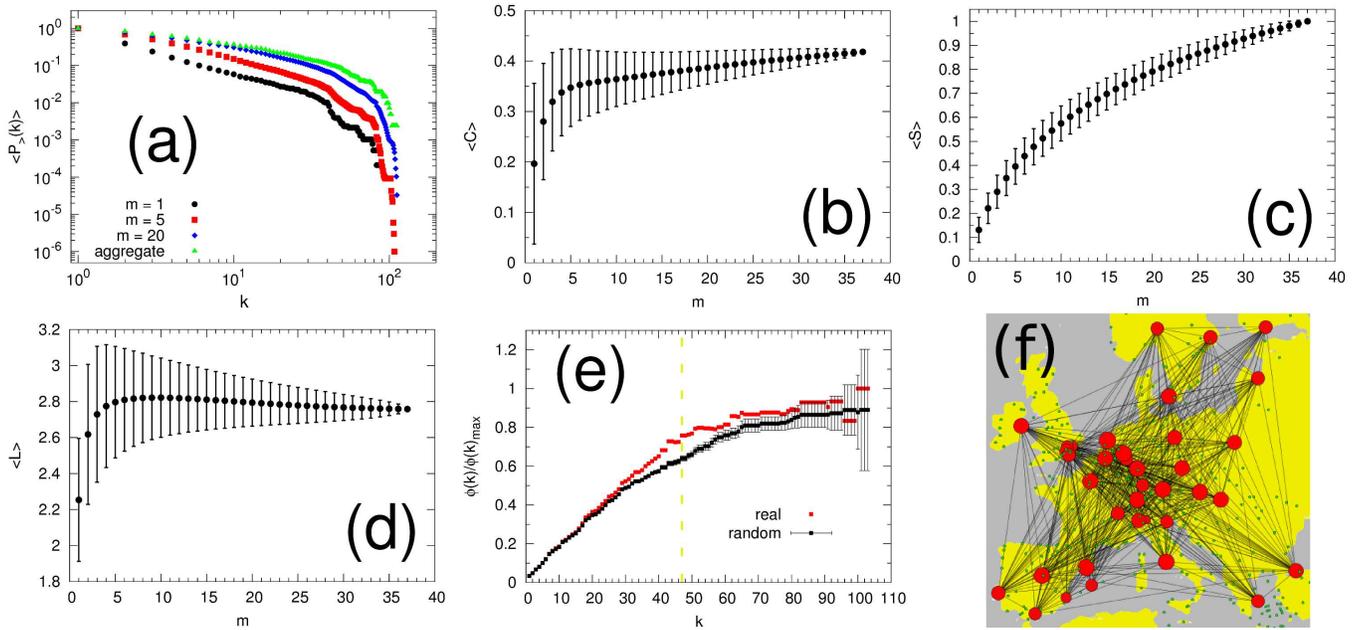}
\caption{Evolution of topological properties of the complete ATN network. (a)
 Average cumulative degree distribution $P_>(k)$ for groups of layers
 merged together: single layers ($\bullet$), five layers
 ($\blacksquare$), twenty layers ($\blacklozenge$) and the aggregate
 ($\blacktriangle$). (b-c-d) Average clustering $\langle C \rangle$,
 size of giant component $\langle S \rangle$, path length $\langle L
 \rangle$ as a function of $m$. (e)
 Link abundance for nodes of degree $k$ or greater, $\phi(k)$ divided
 by its maximum $\phi(k)_{max}$ for the aggregate network in both real
 case ($\blacksquare$) and its randomized version ($\bullet$). The
 vertical dashed line represent the value of $k$ at which the
 difference among the two curves is maximal. (f) A subset of the
 aggregate network showing the connections among those nodes whose
 degree is greater than (or equal to) 47. The size of the nodes is
 proportional to the degree.
 }
\label{fig:results-bigger}
\end{figure*}

We now analyze the evolution of the former measures as more and more layers are merged (independently of whether they do correspond to major or low-cost companies), until the complete aggregate ATN, comprising all the available layers, is reached (see the Methods section for the details on the layer merging procedure).
The results are shown in Fig.~2.

In panel {\bf (a)} we show the evolution for the cumulative degree distribution of the aggregate ATN and those networks obtained by merging $1$, $5$ and $20$ randomly chosen layers. Since right-skewed distributions often display high noise levels at the end of their tails due to the lack of statistics, it is convenient to consider the cumulative distribution instead of the distribution $P(k)$ itself \cite{newman}. A power-law behavior $P_>(k) \propto k^{-\alpha}$ is observed in all the situations considered, with a decrease in the exponent $\alpha$, ranging from $\alpha=1.84$ in the single layer case ($m=1$) to $\alpha=1.39$ for the aggregate ATN. The increase in heterogeneity with the number of layers considered points to a richer-gets-richer phenomenon different from the one seen in classical models for growing scale-free networks: while in the latter case, it results from the addition of new nodes, in the present case it emerges from the addition new layers.

In panel {\bf (b)} we report the clustering coefficient. In this case, we show the behavior of $\langle C \rangle$ as a function of the number of layers used to construct the aggregate ATN,  averaged over the number of different combinations of $m$ elements ($m = 1, \ldots, M$). Interestingly, we see how the clustering suddenly increases as we merge just a few layers: to achieve more than $80\%$ of the final clustering value, we only need to randomly merge together five layers. This result indicates that the large density of triangles present in the ATN is a consequence of the merging of different layers rather than a single-layer property. Thus, in order to make round trips of length $3$ one should make use, most of the times, of more than one airline.

The former result contrasts with the picture obtained for the evolution of the  size of the giant component $\langle S \rangle$. Panel {\bf (c)} describes a monotonous and progressive increase of the {\it coverage} as more layers are aggregated. In fact, around $40\%$ of the European cities are covered when merging together five randomly chosen layers. It is worth noticing that $\langle S \rangle$ also tells us that we are considering a system which is already above the percolation threshold, so that every step towards the aggregate network produces an increment in the collection of reachable destinations (see the value of $\langle S\rangle$ for $m=1$). However, the behavior of the transition for the average path length $\langle L \rangle$ (restricted to those nodes in the giant component) in panel {\bf (d)} shows a rise-and-fall behavior indicating that combining few layers results in the merging of unconnected components at the aggregate level, causing a fast increase in its length. On the other hand, 
after the maximum for $L$ is reached, the addition of new layers has a twofold effect on the giant components: it incorporates new nodes, but also creates alternative links between already present nodes. Thus, the average path length of the giant component balances the addition of new destinations with the creation of new links, and suffers a slow decrease when increasing $m$.

Finally, panel {\bf (e)} shows, only for the aggregate network, the existence of a Rich-club effect quantifying the abundance of links between nodes with degree larger or equal to $k$, $\phi(k)$, normalized with respect to its maximum. This quantity is computed both for the real ATN and for a set of randomized versions of the network in which all the links are rewired keeping the same degree sequence of the original network. This randomization aims at destroying any kind of correlation between the local properties of connected nodes. From the figure it is clear that initially the two curves coincide indicating that the existence of flights between airports with few connections (less than $k=30$) is equally probable in the ATN and in its randomized version. Instead, for $k\in[30,60]$ the points corresponding to the real ATN stand above those corresponding to the randomized network. This result points out that the aggregate ATN displays Rich-club effect (the largest effect being found for $k=47$), thus 
confirming for the European case the findings of Colizza {\it et al.}\cite{
colizza} for the ATN. The existence of such 
effect is quite logical, as usually highly connected nodes correspond to the principal airports of the main European cities which, in most of the cases, are connected among themselves via direct flights. Finally, for $k > 60$ the fluctuations of the randomized case are too large for any statement to be made on the existence of a Rich-club effect.

\subsection{Major {\it versus} Low-cost layers}

\begin{figure*}[ht!]
\centering
\includegraphics[width=\textwidth]{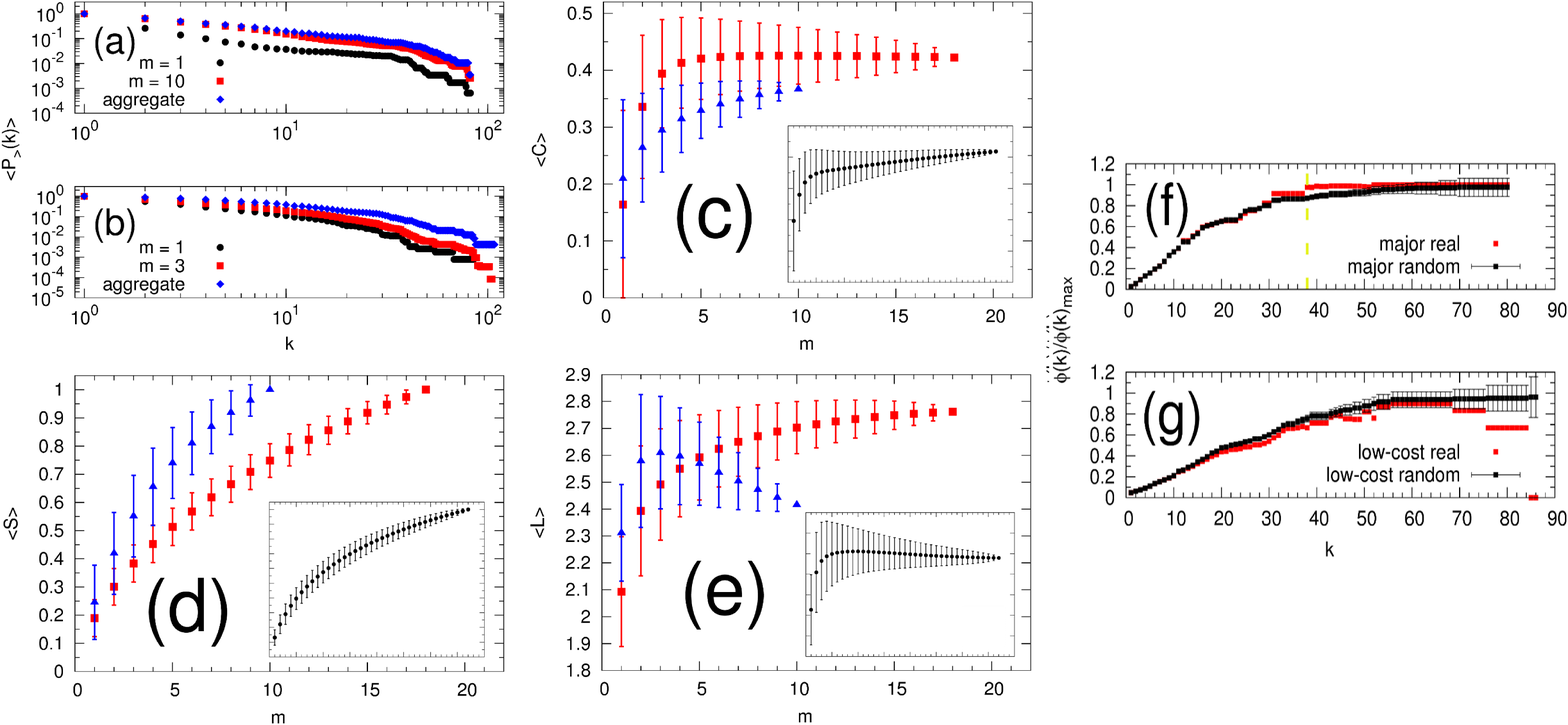}
\caption{Evolution of topological properties of major ($\blacksquare$) and
 low-cost ($\blacktriangle$) subsets. (a-b) Average cumulative degree
 distribution $P_>(k)$ for different number of layers merged together.
 (c-d-e-f) Average clustering $\langle C \rangle$, number of triangles $\langle n_{3}\rangle$, size of giant
 component $\langle S \rangle$, path length $\langle L \rangle$ as a
 function of the number of layers merged. The insets display
 the same quantities in the case of the complete set. (g-h) link
 abundance for the aggregate network. The vertical dashed line
 represents the value of $k$ at which the difference among the two
 curves is maximal.}
\label{fig:results-maj-low}
\end{figure*}

The  European ATN is composed of layers corresponding to airlines of different types. In particular, we find among them {\it major} (national, such as Lufthansa), {\it low-cost} fares (such as Easyjet), {\it regional} (such as Norwegian Air Shuttle) or {\it cargo} (such as Fed-Ex) airlines. These kinds of airlines have developed according to different structural/commercial constraints. For instance, it is known that major airlines are designed following the so-called {\it hub and spoke} structure, to provide an almost complete coverage of the airports belonging to a given country\cite{barthelemy, bryan} and maximize efficiency in terms of national transportation interests. Low-cost companies, instead, tend to avoid overly centralized structures and, to be more competitive, typically cover more than one country simultaneously. To unveil the role that each type of airline plays in the emergence of the topological features of the aggregate ATN, we considered two subsets of layers respectively comprising only {\
it majors} and {\it low-cost} airlines. The results of this study are shown in Fig.~3.

We first address the cumulative degree distribution $P_>(k)$. In the two panels {\bf (a)} and {\bf (b)} we show the distributions $P_>(k)$ for major {\bf (a)} and low-cost {\bf (b)} layers when considering different levels for the merging of the layers of the same kind. For major airlines, the typical trend of a single layer ($m=1$) displays a plateau for moderate values of $k$, indicating a centralized character of this kind of layers, with few hubs having remarkably higher than average connectivity. In addition, when merging more layers ($m=10$ or all the major airlines) the trend shows a rather continuous decay due to the combination of hubs of different size (depending on the nation of the airline). Notice that a hub of a single layer (a single national airline) is highly connected within the same country, but also has some flights to capitals of other European countries which, in turn, are hubs of their corresponding major layers. On the other hand, the cumulative distribution of typical low-cost 
airlines shows a rather different pattern, as its decay is rather progressive, and airports of different size coexist within the same layer.

The differences in organization of low-cost and major airlines is further highlighted by the behavior of the clustering coefficient $\langle C \rangle$. Panel {\bf (c)} shows how major airlines display sharp increases in $\langle C \rangle$ as more major layers are merged, followed by a plateau for $m>5$. This saturation of $C$ is due to the fact that, when merging major layers randomly, national hubs tend to connect together (we have already discussed this fact when introducing the Rich-club effect) in the aggregate network. The saturation of clustering is, however, not observed for the aggregate ATN [see Fig. 2.{\bf (b)} or the inset in panel {\bf (c)}] for which $C(m)$ always increases. This is due to the fact that the merging of low-cost layers leads to a continuous formation of new triangles, thus increasing the clustering with $m$. In addition, in panel {\bf (d)} we show the evolution with $m$ of the average number of triangles, $\langle n_3\rangle$, normalized with respect to the total number of 
triads in the aggregate network for both major and low-cost layers. Interestingly, the monotonic growth of $\langle n_3 \rangle$ reveals that the saturation of the clustering coefficient when $m=5$ for major layers is not due to the fact that new triangles are not added when $m>5$ but to a balance between the new triads and the new connections added when merging additional layers.

The behavior of the giant component $\langle S \rangle$, normalized with respect to the total number of destinations covered by each kind of airline (see panel {\bf (e)}), does not give any particular insight in terms of differences between low-cost and major airlines, except for the fact that in the low-cost case we observe larger fluctuations, mainly due to the large variability in size of the giant component of single layers. On the other hand, the picture described by the average path length $\langle L \rangle$ in panel {\bf (f)} is very interesting. Major and low-cost subsets behave rather differently not only between them, but also with respect to the evolution of the complete set (see inset). For layers corresponding to major companies, $\langle L \rangle$ increases with the number of merged layers. The interpretation of this continuous growth is straightforward: each time a layer corresponding to a major airline is added, even if it shares some common destinations (say some European capitals having 
their corresponding major airlines within the original set of merged layers), the number of new available nodes (small destinations only available through the new added major layer) is large enough to generate an increase in $L$. On the contrary, the case of low-cost displays a rise-and-fall in the behavior of $\langle L\rangle$, due to the large coverage of European countries/cities that already each single low-cost layer displays. Thus, as we merge some of them together, they already cover nearly all the low-cost destinations, and merging of additional layers just adds new connections between them. When combined into the original ATN, these two different trends lead to the saturated evolution of $\langle L \rangle(m)$ shown in the inset.

Finally, we examine once again the onset of the Rich-club effect. From panels {\bf (g)} and {\bf (h)} we notice how the graph corresponding to the aggregate network constructed by merging layers corresponding to major airlines {\bf (g)} displays the presence of a rich club for $k=38$ (almost the same value as in the case of the total aggregate ATN). Interestingly, the Rich-club effect is absent when merging low-cost layers so that, while in the case of major airlines the merge of layers containing large hubs ends up in a system composed of a connected core of  highly connected nodes, the more distributed nature of the low-cost layers prevents the formation of a Rich-club. Thus, a relevant conclusion is that the well-known\cite{colizza} Rich-club effect observed in ATNs is {\it exclusively} related to the presence of major airlines.

\section*{Discussion}

The characterization of the interaction patterns in large systems has recently been spurred by the incorporation of the paradigm of multiplexity. Taking advantage of the European ATN data set, with details of the airlines operating each flight, we showed that the topological properties of the ATN are generally not present in single layers, rather they are the consequence of an emerging phenomenon intimately related to the multilayer character of the system. We also pointed out that the merging of  low-cost and major (national) layers leads to the emergence of qualitatively different aggregate networks. Finally, we demonstrated that the combination of these two different behaviors accounts for the many important structural features of the global ATN, such as the Rich-club effect (mainly due to the layers of major airlines), path redundancy (resulting from a cooperative combination of the clustering of low-cost and major layers), or small-worldness (remarkably enhanced by the presence low-cost layers).

Our study highlights the importance of considering the multiplex character of most real networked systems, and shows that considering layers as relevant entities of a network (such as nodes and links at the micro-scale or communities at the meso-scale) will contribute to a better understanding and modeling of dynamical processes taking place at the level of aggregate network.

\appendix


\section{Dataset}
The data analyzed in this paper are taken from the complete list of airlines operating Instrumental Flight Rules (IFR) flights between European airports on a certain day obtained from EUROCONTROL and the Complex World Network in the context of the SESAR Work Package~E \cite{complexworld}. We selected only those airlines whose number of {\it destinations} is above the average (which is 32), obtaining $\mathcal{L} = 37$ different airlines (layers), that include both {\it major} companies (like Lufthansa or Air France), and {\it low-fares} (low-cost) companies (as Ryanair or Easyjet). Each layer $\ell$ in this multiplex representation is a graph $G^\ell=(\mathcal{N}^\ell,\mathcal{E}^\ell)=(\mathcal{N},\mathcal{E}^\ell)$ with $\mathcal{N}^\ell=\mathcal{N}=450$ nodes and $K^\ell$ links that models a single airline. An example of such networks is shown in Fig.~1. The ensemble of all these layers constitutes our multilayer system, that we will call the {\it complete set}. We will also consider the subset of {\it 
major} airlines, that will be a multiplex network made of $\mathcal{L}^\prime = 18$ layers,   and the subset of {\it low-cost} companies, with $\mathcal{L}^{\prime\prime} = 10$ layers. Note that the remaining airlines, such as cargo airlines, constitutes a marginal small subset and therefore its analysis is residual.

\section{Topological indexes}
In this section, we  present a summary of the topological measures used throughout the paper. Note that the considered topological measures are essentially defined for classic monoplex  networks, and their extensions to the multiplex setting is an exercise, whose details are here shown.

One of the most basic topological parameter of a complex network $G=(\mathcal{N},\mathcal{E})$  is the {\it degree distribution} $P(k)$ which is defined as the probability that a node chosen uniformly at random has degree $k$, or equivalently the fraction of nodes in the network having degree $k$ \cite{boccaletti, newman}. Since broad distributions often display high noise levels at the end of their tails, here related to the low abundance of highly connected nodes, it is convenient to consider the {\it cumulative distribution} $P_>(k)$. Cumulative distribution $P_>(k)$ is the probability that a randomly chosen node has a degree equal or greater than $k$, i.e.
\begin{equation}
\label{eq:pk}
P_>(k) = \frac{1}{N} \sum_{k^\prime = k}^{\infty} N(k) \,,
\end{equation}
where $N(k)$ is the number of nodes with degree $k$ and $N=|\mathcal{N}|$ is the total number of nodes in the network.

The {\it average path length} \cite{watts} $L(G)$ is the average length of the shortest paths among all the couples of nodes in the network, i.e.
\begin{equation}
\label{eq:avg_len}
L(G) = \frac{1}{N (N -1)} \sum_{i,j \in \mathcal{N}} d_{ij} \,,
\end{equation}
where $d_{ij}$ is the minimum number of hops one has to make to go from node $i$ to node $j$ in $G$ (the {\it distance} from $i$ to $j$). Note that this definition diverges if $G$ is not connected, since $d_{ij}$ may be infinite. One way to avoid this divergence is considering the average only on the largest connected component, and an alternative approach that has been shown very useful in many cases is considering the harmonic mean of the distances.

The (local) {\it clustering coefficient} \cite{watts}  $c_i$ of a node $i\in \mathcal{N}$ is defined as
\begin{equation}
\label{eq:clustering}
c_i = \frac{2 \, e_i}{k_i (k_i - 1)} \,,
\end{equation}
where $e_i$ is the number of neighbors of $i$ which are mutual neighbors, and $k_i$ is the degree of node $i$. Therefore the (local) clustering coefficient of a node $i$ is the ratio between the number of neighbors of $i$ which are mutual neighbors and the maximal possible number of edges between  neighbors of $i$.  The (average) clustering coefficient $C$ of a graph is the arithmetic mean of $c_i$ over all its nodes.

The {\it giant component}  $S(G)$ is the largest connected component of $G$ and the {\it size of the giant component} is the proportion of nodes in the network that belong to the giant component, i.e.,
\begin{equation}
\label{eq:giant}
S(G) = \max_{i \in \mathcal{N}}\frac{N_i}{N}
\end{equation}
where $N_i$ is the number of nodes of the maximal connected subnetwork of $G$ containing node $i$.

If we take a node with {\it degree $0\le k\le |\mathcal{N}|$}, the {\it Rich-club coefficient} $R(k)$\cite{zhou} is given by
\begin{widetext}
\centering
\begin{equation}
\label{eq:rich-club-coeff}
R(k) = \frac{\phi(k)}{\phi(k)_{max}} \left(\frac{\phi^\prime(k)}{\phi^\prime(k)_{max}}\right)^{-1} = \frac{2 \phi(k)}{N_{>k}(N_{>k} - 1)} \frac{N_{>k}(N_{>k} - 1)}{2 \phi^\prime(k)} = \frac{\phi(k)}{\phi^\prime(k)},
\end{equation}
\end{widetext}
where
\begin{itemize}
\item[{\it(i)}] $\phi(k)$ is the number of edges connecting nodes of degree greater or equal to $k$  (called the {\it link abundance}),
\item[{\it(ii)}] $\phi(k)_{max}$ is the maximum number of links that can exist between nodes of degree $k$,
\item[{\it(iii)}] $\phi'(k)$ is the {\it link abundance} on a network with the same degree sequence of the original but with connections randomly shuffled.
\item[{\it (iv)}] $\phi'(k)_{max}$ is the maximum number of links that can exist between nodes of degree $k$ on a network with the same degree sequence of the original but with connections randomly shuffled.
\item[{\it(v)}] $N_{>k}$ is the number of nodes with degree greater or equal to $k$.
\end{itemize}
If, for a certain value of $k$, $R(k)>1$ for some $0\le k\le N=|\mathcal{N}|$, then we say that $G$ has a {\it Rich-club}. Note that in the plots presented in this paper, we decided to present the ratios  $\phi(k) / \phi(k)_{max}$ and $\phi'(k) / \phi'(k)_{max}$ instead of $R(k)$. 
The randomization, in our case, is repeated 1,000 times, while the shuffling is repeated 10,000 times to ensure a robust statistical sampling. Note that for the ATN network, having a size of $N=450$ nodes, the number of random shuffling steps is large enough to guarantee that the resulting network is fully randomized. This randomization method is known as {\it Markov Chain Monte Carlo Algorithm} \cite{blitzstein11}. However, for bigger graphs other methods are recommended so to minimize the computation cost for producing reliable randomized networks, see the work by Del Genio {\it et al.} \cite{delgenio10}.

Next, we describe the {\it layer merging} procedure used to study the evolution of the topological measures and the behavior of the layers in the major airline and low-cost multiplex sub-networks.

If we fix a subset of layers  $\{G^\ell;\enspace \ell=\ell_1,\cdots,\ell_m\}$ to merge together, we construct a monoplex network $G'=(\mathcal{N},\mathcal{E}')$ (i.e. a classic network with only one layer) given by
\[
G^\prime= \bigcup_{j=1}^m G^{\ell_j}.
\]
This network $G'$ is obtained by projecting all the $m$ layers onto one and by converting multiple links into single ones.

Now if we fix $m$, we look for all the possible mergings of $m$ layers, The number of different configurations to arrange $n$ layers into groups of size $m$ without repetitions is given by $C^{n}_{m} =  {n \choose m}$, therefore if we want to compute a topological measure on the ensemble of $m$ layers, we should first compute it on each of the $C^n_m$ mergings, and then average over all $C^n_m$ possible configurations.  However, when the number of possible configurations exceeded a certain threshold, we operated a random sampling over 500,000 mergings in order to avoid the growth of the computation time. Throughout the paper the operator $\langle \cdot \rangle$ denotes the average over the elements of the ensemble.  As an example, if we want to compute the clustering coefficient over an ensemble $\mathcal{C}$, we compute:
\begin{equation}
\label{eq:avg_clust}
\langle C \rangle = \frac{1}{N_{comb}} \sum_{i \in \mathcal{C}} C_i \,,
\end{equation}
where $N_{comb}$ is the number of elements of $\mathcal{C}$ and $C_i$ is the average clustering of the network obtained merging together the layers corresponding to $i\in\mathcal{C}$.

\section*{Acknowledgments}

This work has been partially supported by the Spanish MINECO under projects MTM2009-13848 and FIS2011-25167; European FET project MULTIPLEX (317532); and by the Comunidad de Arag\'on through Project No. 2009SGR0838, FMI22/10. J.G.G is supported by MINECO through the Ram\'on y Cajal program.




\end{document}